\documentclass[showpacs,amsmath,amssymb,superscriptaddress]{revtex4-1}
\usepackage{dcolumn}
\usepackage{color}
\usepackage{graphicx}
\usepackage{multirow}
\usepackage{amsmath}
\usepackage{bm}

\begin{document}

\title{Signatures of shape phase
transitions in odd-mass nuclei}

\author{K.~Nomura}
\affiliation{Physics Department, Faculty of Science, University of
Zagreb, 10000 Zagreb, Croatia}
\affiliation{Center for Computational Sciences, University of Tsukuba,
Tsukuba 305-8577, Japan}
\author{T.~Nik\v si\'c}
\author{D.~Vretenar}
\affiliation{Physics Department, Faculty of Science, University of Zagreb, 10000 Zagreb, Croatia}

\date{\today}

\begin{abstract}

Quantum phase transitions between competing ground-state shapes of atomic nuclei with 
an odd number of protons or neutrons are investigated in a microscopic
 framework based on nuclear energy density functional theory and the
 particle-plus-boson-core coupling scheme. 
The boson-core Hamiltonian, as well as the single-particle energies and occupation
 probabilities of the unpaired nucleon, are completely determined by constrained 
 self-consistent mean-field calculations for a specific choice of the
 energy density functional and paring interaction, and only the strength
 parameters of the particle-core coupling are adjusted to reproduce selected 
 spectroscopic properties of the odd-mass system.   
We apply this method to odd-$A$ Eu and Sm isotopes with neutron number $N \approx 90$, 
and explore the influence of the single unpaired fermion on the occurrence of a shape phase 
transition. Collective wave functions of low-energy states are used to compute
quantities that can be related to quantum order parameters: deformations, 
excitation energies, E2 transition rates and separation energies, and their evolution with  
the control parameter (neutron number) is analysed. 
\end{abstract}



\maketitle


\section{Introduction}

Quantum phase transitions (QPTs) are a prominent feature of many-body systems in 
many fields of physics and chemistry \cite{carr-book}. 
Nuclear QPTs \cite{cejnar2010} are transitions between competing 
ground-state shapes (spherical, axially deformed, shapes that are soft 
with respect to triaxial deformations) induced by variation of a non-thermal control 
parameter at zero temperature. Gradual transitions between different shapes in chains of isotopes or 
isotones predominate but in a number of cases, with the addition or subtraction of only 
few nucleons, abrupt changes in ground-state properties are 
observed and related critical phenomena emerge \cite{BM,CasBook}. 
When considering QPT in finite systems such as atomic nuclei, in particular, an
essential question is how to identify observables that can be related to order parameters. 
In addition, discontinuities at a phase transitional point are smoothed out in finite nuclei, 
and it is not always possible to associate the point of phase transition with a particular nucleus, 
because the control parameter of shape phase transitions, that is, the nucleon number, is not 
continuous. Numerous experimental studies of transitional nuclei have been carried out 
in the last fifteen years, and signatures of first- and second-order QPTs have been identified and 
investigated with various theoretical methods (for a review see Ref.~\cite{cejnar2010} and references 
cited therein). New and very active areas of research include  
excited-state quantum phase transitions
\cite{iachello2011a,stransky2014,stransky2015,cejnar2015}, and QPTs in odd-mass nuclei
\cite{iachello2011,iachello2011a,petrellis2011}.

QPTs between equilibrium shapes of even-even nuclei, that is, systems with both 
proton and neutron numbers ($Z$ and $N$) even, have been extensively explored  
using a variety of phenomenological \cite{cejnar2010} and microscopic 
\cite{niksic2007,robledo2008,li2009b,nomura2013oct} approaches. 
A description of possible QPTs in odd-mass nuclei, in which either $Z$ or $N$ is an odd number, 
is considerably more complex. Because of the effect of pairing, 
in even-even systems all nucleons are coupled pairwise to $T=1$ pairs, and the
low-energy excitation spectra are characterised by collective
vibrational and rotational degrees of freedom \cite{BM}. 
For odd-$A$ nuclei both single-particle
(unpaired fermion(s)) and collective (even-even 
core) degrees of freedom determine the low-energy excitations \cite{bohr1953}. 
Important issues when considering QPTs in odd-mass nuclei are the influence of the 
unpaired fermion(s) on the location and nature of the phase transition, empirical 
signatures of QPTs in odd-$A$ nuclei, and the definition and computation of 
order parameters \cite{iachello2011,petrellis2011}. To address these questions 
shape phase transitions in odd-mass systems have mainly been investigated  
using empirical approaches such as 
algebraic methods \cite{boyukata2010,iachello2011,petrellis2011,boyukata2015}, and geometrical 
 models \cite{zhang2013a,zhang2013b}. Microscopic 
studies of QPTs in odd-mass nuclei, and in particular 
studies of quantum order parameters, have not been as 
extensively pursued as in the the case of even-even systems.

This work presents a microscopic study of nuclear shape phase
transitions in odd-mass systems in the rare earth region with $N\approx 90$ and, in particular, 
an analysis of observables that can be related to order parameters.  
Recently we have developed a new theoretical method \cite{nomura2016odd} based on nuclear density
functional theory \cite{bender2003,vretenar2005,niksic2011} and the
particle-core coupling scheme \cite{bohr1953,BM}. 
The even-even core nucleus is modelled in terms of $s$ and $d$ boson 
degrees of freedom \cite{IBM}, which represent correlated pairs of valence nucleons, 
and the particle-core coupling of the unpaired proton or neutron 
is described in the framework of the interacting
boson-fermion model (IBFM) \cite{IBFM}. 
In the model of Ref.~\cite{nomura2016odd} the parameters of the even-even boson core
Hamiltonian, and the single-particle energies and occupation probabilities of
the odd-fermion states, are completely determined by a constrained self-consistent
mean-field (SCMF) calculation for a
specific choice of the energy density functional (EDF) and pairing interaction.
Only the strength parameters of the fermion-boson coupling terms of the Hamiltonian are
specifically adjusted to reproduce selected spectroscopic data for a given nucleus. 
The method has been illustrated and tested in an analysis of axially-deformed 
odd-$A$ nuclei $^{151-155}$Eu, and it has been shown that the approach enables a systematic, accurate and
computationally feasible description of low-energy spectroscopic properties of odd-mass nuclei \cite{nomura2016odd}.

In the present analysis we consider the structural evolution of odd-mass Eu ($Z=63$) and Sm
($Z=62$) isotopes in the region with neutron number $N\approx 90$. For the low-energy 
excitation spectra these systems can be treated as a single   
unpaired proton and neutron, respectively, coupled to the even-even core Sm nuclei. 
The boson core nuclei, $^{146-154}$Sm, provide an outstanding 
example of a first-order QPT from spherical to axially-deformed equilibrium
shapes, with the control parameter being the neutron number \cite{cejnar2010,X5}. 
The odd particle for the odd-$A$ Sm nuclei is also a neutron, while that of the odd Eu isotopes 
is a proton, and this means a nucleon different from the control parameter of the boson
core. We analyse the influence of the unpaired nucleon on the occurrence of a QPT in both cases.

Section \ref{sec:method} contains a concise outline of the theoretical method used in the present study. 
In Sec. \ref{sec:qpt} we analyse the theoretical low-energy positive- and negative-parity excitation 
spectra of odd-$A$ Sm nuclei in comparison with available data, and explore signatures of spherical 
to axially-deformed shape transitions in odd-mass Eu ($Z=63$) and Sm ($Z=62$): equilibrium 
deformation parameters, spectroscopic properties, and separation energies.  
Section \ref{sec:conclusion} includes a summary and a brief outlook for future studies. 

\section{Model for odd-mass nuclei\label{sec:method}}

In Ref.~\cite{nomura2016odd} we introduced a novel method for calculating spectroscopic properties of 
medium-mass and heavy atomic nuclei with an odd number of nucleons, based on the framework of 
nuclear energy density functional theory and the particle-core coupling scheme.
The model Hamiltonian $\hat H$ used to describe an odd-$A$ nucleus contains a term that
corresponds to the even-even boson core $\hat H_B$ (built from monopole
$s$ (with spin and parity $J^{\pi}=0^+$) and quadrupole $d$
($J^{\pi}=2^+$) bosons), a single-particle
Hamiltonian $\hat H_F$ that describes the unpaired nucleon(s), and an
interaction term $\hat H_{BF}$ that couples the boson and fermion degrees of freedom: 
\begin{eqnarray}
\label{eq:ham}
 \hat H=\hat H_B + \hat H_F + \hat H_{BF}. 
\end{eqnarray}
The number of bosons $N_B$ and the number of odd fermions $N_F$ are
conserved separately and, since we consider low-energy excitation spectra of odd-mass 
systems, $N_F=1$. In the present version, therefore, the model space does not include 
three and higher (quasi)particle states. 
Since a boson represents a collective pair of valence nucleons, $N_B$ corresponds to
the number of fermion pairs, particle or hole, in the major valence shell \cite{OAI}. 
In the present case $N_B$ equals the number of fermion pairs outside 
the doubly-magic nucleus $^{132}$Sn, that is, from 7 to 11 for the boson
core nuclei $^{146-154}$Sm. 
We employ for the boson-core Hamiltonian the following form: 
\begin{eqnarray}
\label{eq:ibm}
 \hat H_B = \epsilon_d\hat n_d + \kappa\hat Q_B\cdot\hat Q_B +
  \kappa^{\prime}\hat L\cdot\hat L, 
\end{eqnarray}
with the $d$-boson number operator $\hat n_d=d^{\dagger}\cdot\tilde d$,
the quadrupole operator $\hat Q_B=s^{\dagger}\tilde d+d^{\dagger}\tilde s +
\chi[d^{\dagger}\times\tilde d]^{(2)}$, and the angular momentum
operator $\hat L=\sqrt{10}[d^{\dagger}\times\tilde d]^{(1)}$. 
$\epsilon_d$, $\kappa$, $\kappa^{\prime}$ and $\chi$ are parameters. 
The fermion Hamiltonian for a single nucleon reads $\hat
H_F=\sum_{j}\epsilon_j[a^{\dagger}_j\times\tilde a_j]^{(0)}$, 
with $\epsilon_j$ the single-particle energy of the spherical orbital $j$. 
For the particle-core coupling $\hat H_{BF}$ we use the simplest form \cite{IBFM,IBFM-Book}: 
\begin{eqnarray}
\label{eq:bf}
 \hat H_{BF}=\sum_{jj^{\prime}}\Gamma_{jj^{\prime}}\hat
  Q_B\cdot[a^{\dagger}_j\times\tilde a_{j^{\prime}}]^{(2)}
+\sum_{jj^{\prime}j^{\prime\prime}}\Lambda_{jj^{\prime}}^{j^{\prime\prime}}
:[[d^{\dagger}\times\tilde a_{j}]^{(j^{\prime\prime})}
\times
[a^{\dagger}_{j^{\prime}}\times\tilde d]^{(j^{\prime\prime})}]^{(0)}:
+\sum_j A_j[a^{\dagger}\times\tilde a_{j}]^{(0)}\hat n_d, 
\end{eqnarray}
where the first, second and third terms are referred to as the quadrupole dynamical, 
exchange, and monopole interactions, respectively. 
The strength parameters $\Gamma_{jj^{\prime}}$, $\Lambda_{jj^{\prime}}^{j^{\prime\prime}}$ and $A_j$ can be expressed, by use of
the generalized seniority scheme, in the following $j$-dependent forms \cite{scholten1985}: 
\begin{eqnarray}
\label{eq:bf-strength}
&&\Gamma_{jj^{\prime}}=\Gamma_0\gamma_{jj^{\prime}} \\
&&\Lambda_{jj^{\prime}}^{j^{\prime\prime}}=-2\Lambda_0\sqrt{\frac{5}{2j^{\prime\prime}+1}}\beta
_{jj^{\prime}}\beta_{j^{\prime}j^{\prime\prime}} \\
&&A_j=-A_0\sqrt{2j+1}
\end{eqnarray}
where
$\gamma_{jj^{\prime}}=(u_ju_{j^{\prime}}-v_jv_{j^{\prime}})Q_{jj^{\prime}}$
and 
$\beta_{jj^{\prime}}=(u_jv_{j^{\prime}}+v_ju_{j^{\prime}})Q_{jj^{\prime}}$,
and the matrix element of the quadrupole operator in the single-particle
basis $Q_{jj^{\prime}}=\langle j||Y^{(2)}||j^{\prime}\rangle$. 
The factors $u_j$ and $v_j$ denote the occupation probabilities of the
orbit $j$, and satisfy $u_j^2+v_j^2=1$. 
$\Gamma_0$, $\Lambda_0$ and $A_0$ denote the strength parameters. 
A more detailed discussion of each term of the Hamiltonian in Eq.~(\ref{eq:ham})
is included in Ref.~\cite{nomura2016odd}.

To build the boson-fermion Hamiltonian in a first step  
one determines the parameters of the boson Hamiltonian $\hat H_B$, 
following the procedure introduced in Ref.~\cite{nomura2008}: 
the microscopic deformation energy surface, calculated with the constrained
self-consistent mean-field (SCMF) method for a specific choice of the 
nuclear energy density functional (EDF) and a pairing interaction, 
is mapped onto the corresponding expectation value of
the interacting boson Hamiltonian in the boson coherent state \cite{ginocchio1980}. This procedure
uniquely determines the parameters in the boson Hamiltonian $\hat H_B$. 
 
The fermion model space contains all the spherical major shell valence orbitals of the 
unpaired particle, proton or neutron. In the present calculation 
we include spherical single-particle
orbitals in the proton major shell $Z=50-82$ (positive parity $1g_{7/2}, 2d_{5/2},
2d_{3/2}, 3s_{1/2}$, and negative
parity $1h_{11/2}$ ) for the odd-$Z$ Eu isotopes, and the orbitals in the
neutron major shell $N=82-126$ (positive parity $1i_{13/2}$, and
negative parity $1h_{9/2}, 2f_{7/2}, 2f_{5/2}, 3p_{3/2}, 3p_{1/2}$) for the odd-$N$ Sm nuclei. 
The canonical single-particle energies and occupation probabilities of
these orbitals determine the terms $\hat H_F$ and
$\hat H_{BF}$ of the Hamiltonian, respectively, and are obtained from the SCMF 
calculation constrained to zero deformation. 

In the final step the strength parameters of the boson-fermion
Hamiltonian $\hat H_{BF}$ are adjusted for each nucleus separately. 
Optimal values of the corresponding strength
parameters ($\Gamma_0$, $\Lambda_0$ and $A_0$) are adjusted  to reproduce
the ground-state spin and/or the excitation energies of a few lowest
levels, separately for positive- and negative-parity states.

The resulting Hamiltonian of 
Eq.~(\ref{eq:ham}) is diagonalized numerically \cite{PBOS} in the spherical basis
$|j,L,\alpha,J\rangle$, where $\alpha$ is a generic notation for a set of quantum numbers
$n_d,\nu,n_{\Delta}$ that distinguish states with the same 
boson angular momentum $L$ \cite{IBM}, and $J$ is the total angular momentum of
the Bose-Fermi system ($|L-j|\leq J\leq L+j$).  
Using the wave functions resulting from the diagonalization, 
electromagnetic transition rates can be calculated. The relevant
decay mode in the present study is the electric quadrupole (E2)
transition. The E2 transition operator of the Bose-Fermi system reads 
$\hat T^{\rm (E2)}=e_B\hat Q_B + e_F\hat Q_F$, where $\hat Q_B$ and
$\hat Q_F$ are the quadrupole operators for the boson and fermion systems \cite{nomura2016odd},
respectively, and $e_B$ and $e_F$ are the effective charges. 
$e_B$ is adjusted to reproduce the experimental
$B({\rm E2}; 2^+_1\rightarrow 0^+_1)$ value for the boson-core nucleus,
while the constant value $e_F=1.0$ $e$b is used for the fermion effective charge.

\section{Quantum shape phase transitions in odd-mass Sm and Eu isotopes\label{sec:qpt}}

Probably the best example of a QPT in atomic nuclei is in the rare earth region with 
$N \approx 90$ neutrons, where a transition between spherical and axially symmetric 
equilibrium shapes has been extensively investigated both experimentally
\cite{casten1998,zamfir1999,klug2000,kruecken2002,tonev2004,moeller2006,kulp2008}, and by using a 
number of theoretical methods \cite{X5,casten2001,mccutchan2004,maccutchan2005,niksic2007,werner2008,li2009}. 
Sm nuclei in this mass region, and $^{152}$Sm in particular, was the
first reported empirical example of a structure at the critical point of
a first-order transition between a vibrator and the axial rotor phase \cite{casten2001}.
Here we analyse low-energy states of odd-proton Eu nuclei and odd-neutron Sm nuclei that can be 
described by coupling the corresponding unpaired nucleon to the even-even Sm boson core. 

\begin{figure}[htb!]
\begin{center}
\begin{tabular}{cc}
\includegraphics[scale=0.42]{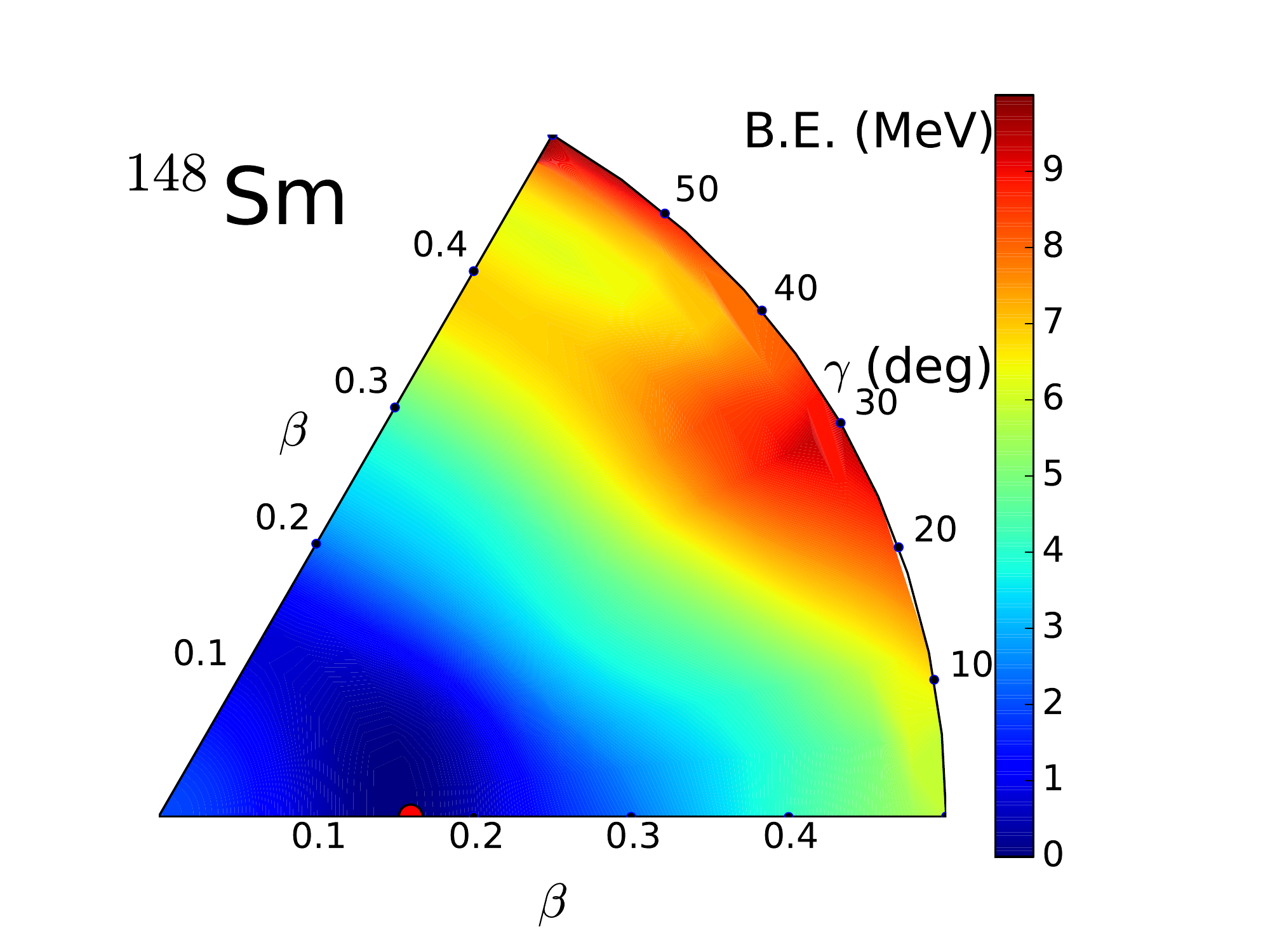}&
\hspace{-1cm}\includegraphics[scale=0.42]{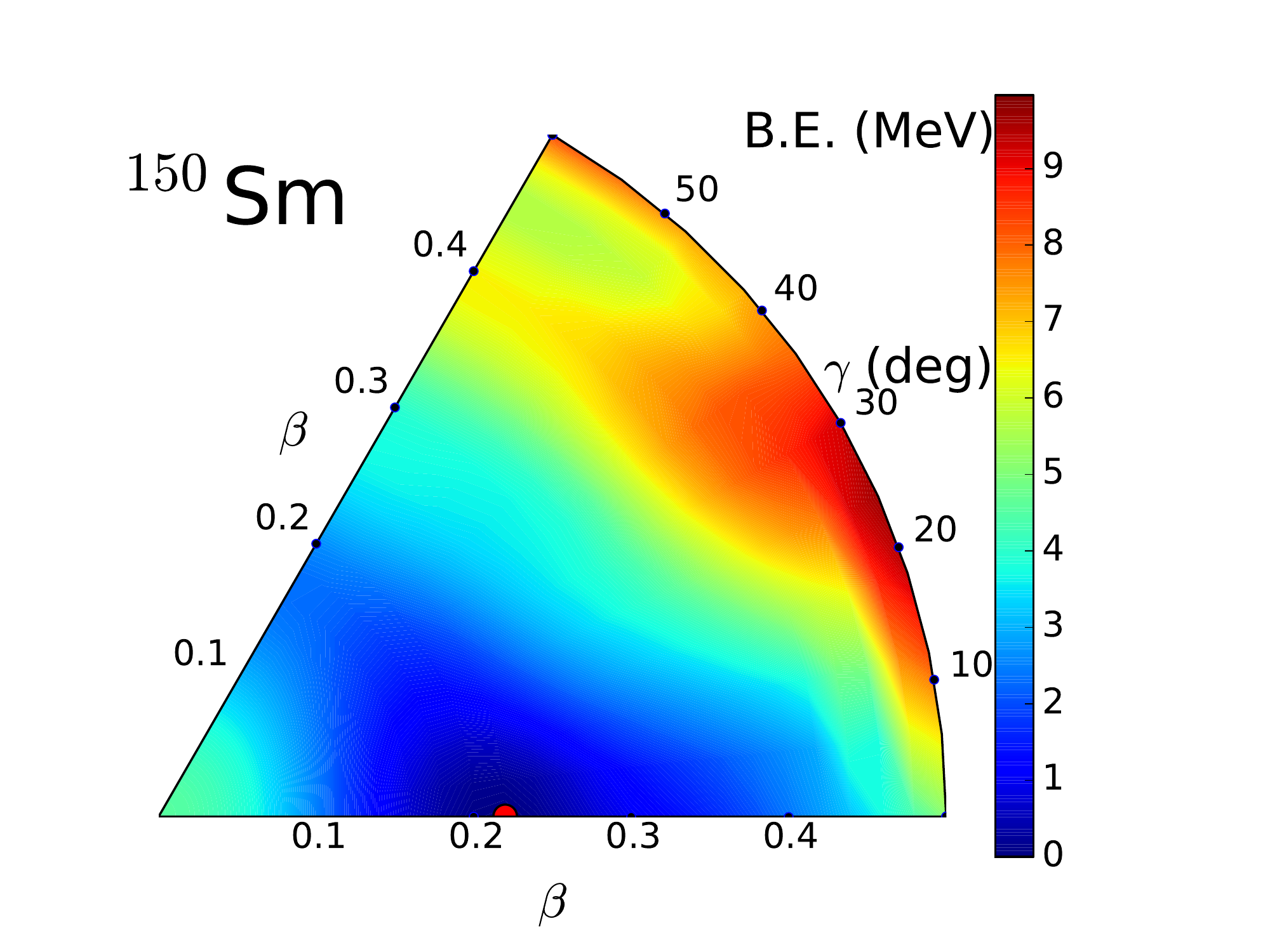} \\
\includegraphics[scale=0.42]{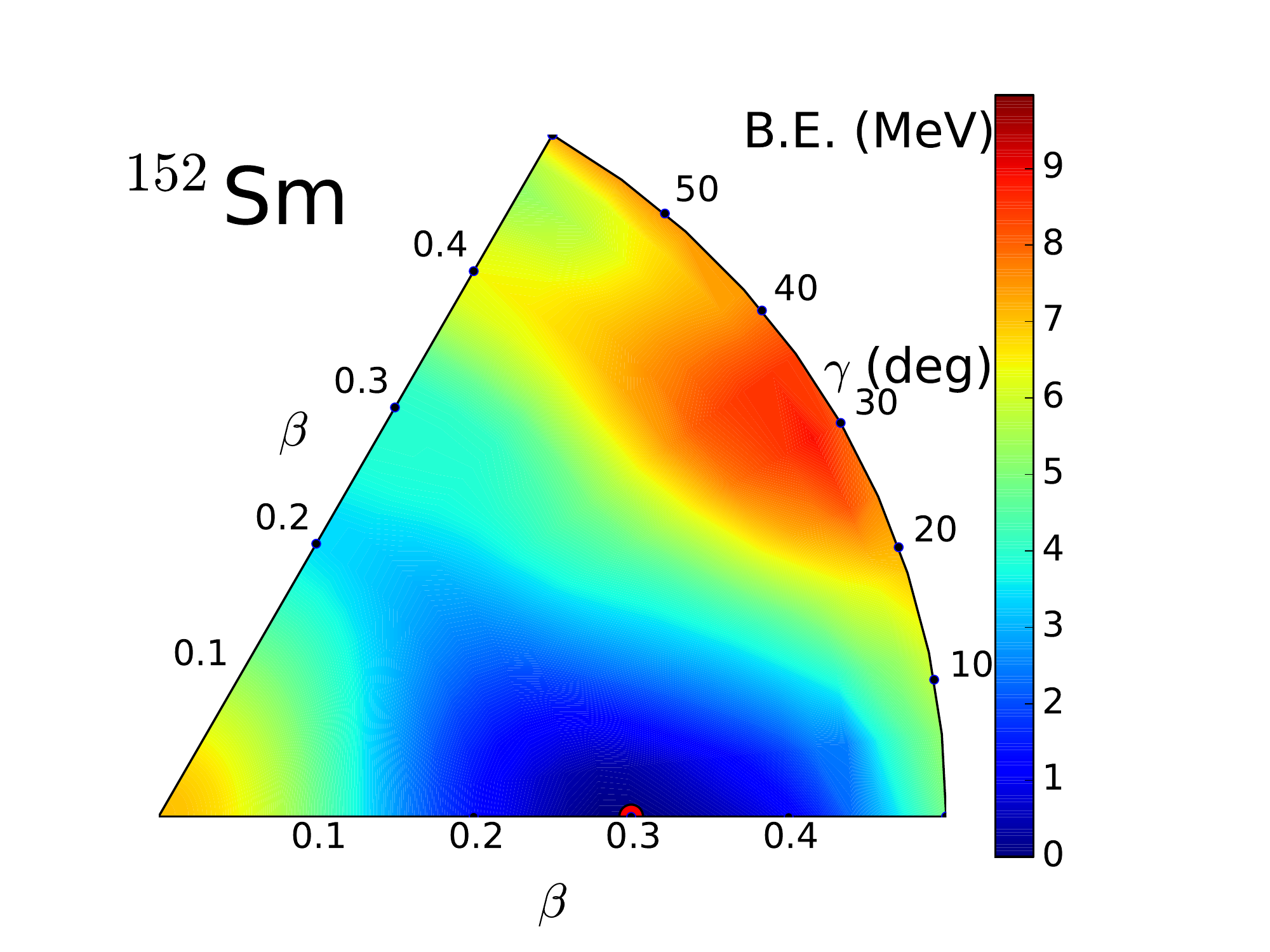}&
\hspace{-1cm}\includegraphics[scale=0.42]{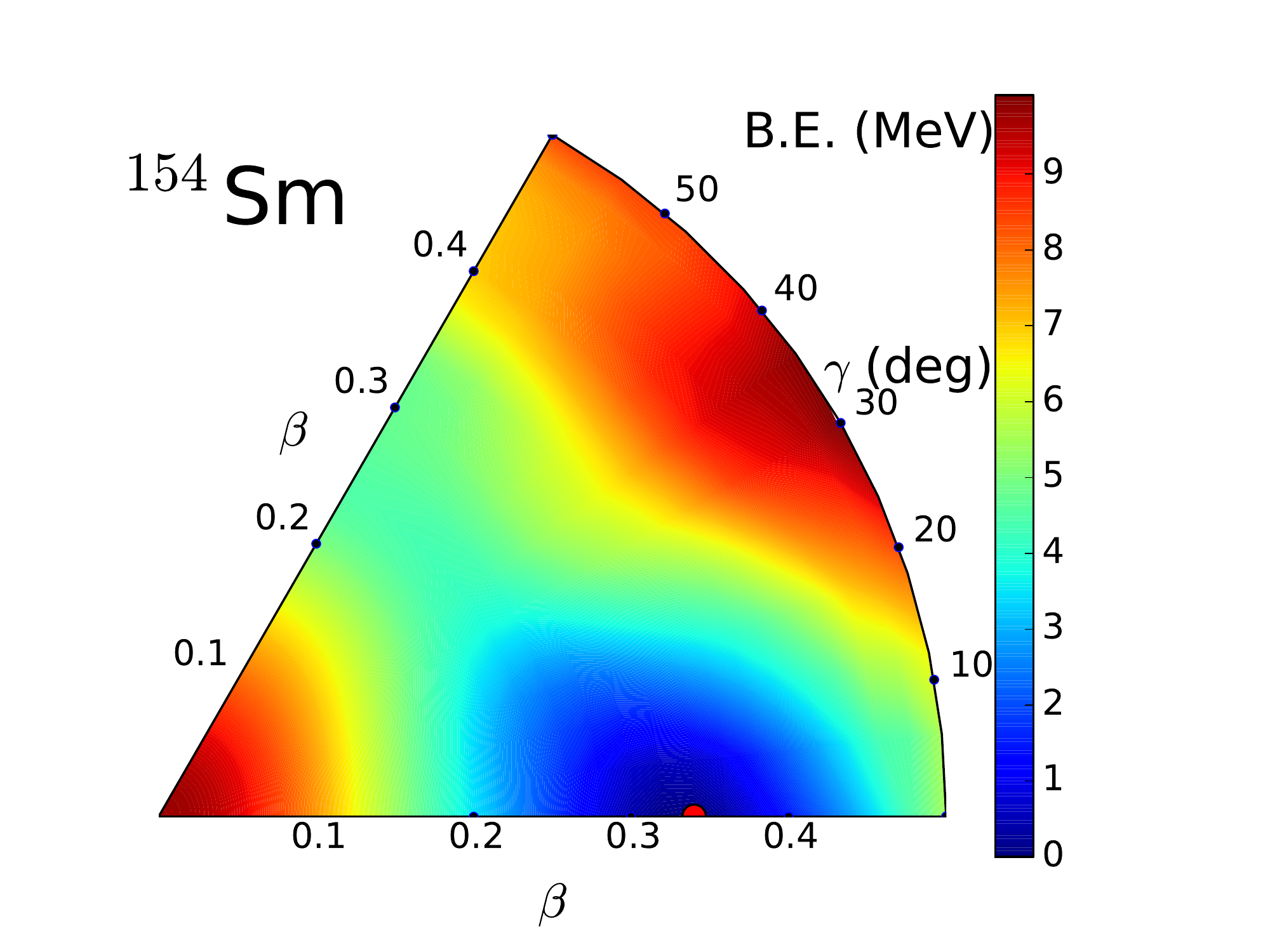} 
\end{tabular}
\caption{(Color online) Self-consistent RHB triaxial quadrupole
binding energy maps of the even-even $^{148-154}$Sm isotopes
in the $\beta - \gamma$ plane ($0\le \gamma\le 60^{\circ}$).
For each nucleus the energy surface 
is normalized with respect to
the binding energy of the absolute minimum.}
\label{fig:pes}
\end{center}
\end{figure}

The deformation energy surfaces for a set of even-even Sm core nuclei, which determine 
the parameters of the interacting-boson Hamiltonian, are calculated as functions of the 
polar deformation parameters $\beta$ and $\gamma$, using the the constrained self-consistent
relativistic Hartree-Bogoliubov (RHB) model based on the energy density functional DD-PC1
\cite{DDPC1}, and a separable pairing force of finite range \cite{tian2009}. 
The map of the energy surface as a function of quadrupole deformation is obtained by imposing 
constraints on the axial and triaxial mass quadrupole moments. 
In Fig.~\ref{fig:pes}  we display the self-consistent
RHB triaxial quadrupole binding energy maps of
the even-even $^{148 -154}$Sm in the $\beta - \gamma$ plane ($0\le \gamma\le 60^{\circ}$),
The energy maps clearly exhibit a 
gradual increase of deformation of the prolate minimum with increasing neutron 
number, from spherical $^{146,148}$Sm to well-deformed prolate shapes at and beyond $^{154}$Sm, 
and the evolution of the $\gamma$-dependence of 
the potentials. The axial potential barrier at zero deformation increases with mass number. 
With increasing $N$ the prolate deformation of Sm isotopes 
at equilibrium becomes more pronounced and the shape evolution corresponds, 
in the language of the interacting boson model, to a 
transition between the U(5) and SU(3) limits of the Casten symmetry 
triangle \cite{IBFM-Book}. The energy surfaces of $^{150,152}$Sm indicate that these are transitional nuclei,
characterised by a softer potential around the equilibrium minimum both in the
 $\beta$ and $\gamma$ directions. 
The softness of the energy surface with respect to the quadrupole 
deformation parameters $\beta$ and/or $\gamma$ around the mean-field equilibrium minimum 
has been associated with the phenomenon of quantum shape phase
transition \cite{cejnar2010}. 

\begin{figure*}[htb!]
\begin{center}
 \includegraphics[width=0.7\linewidth]{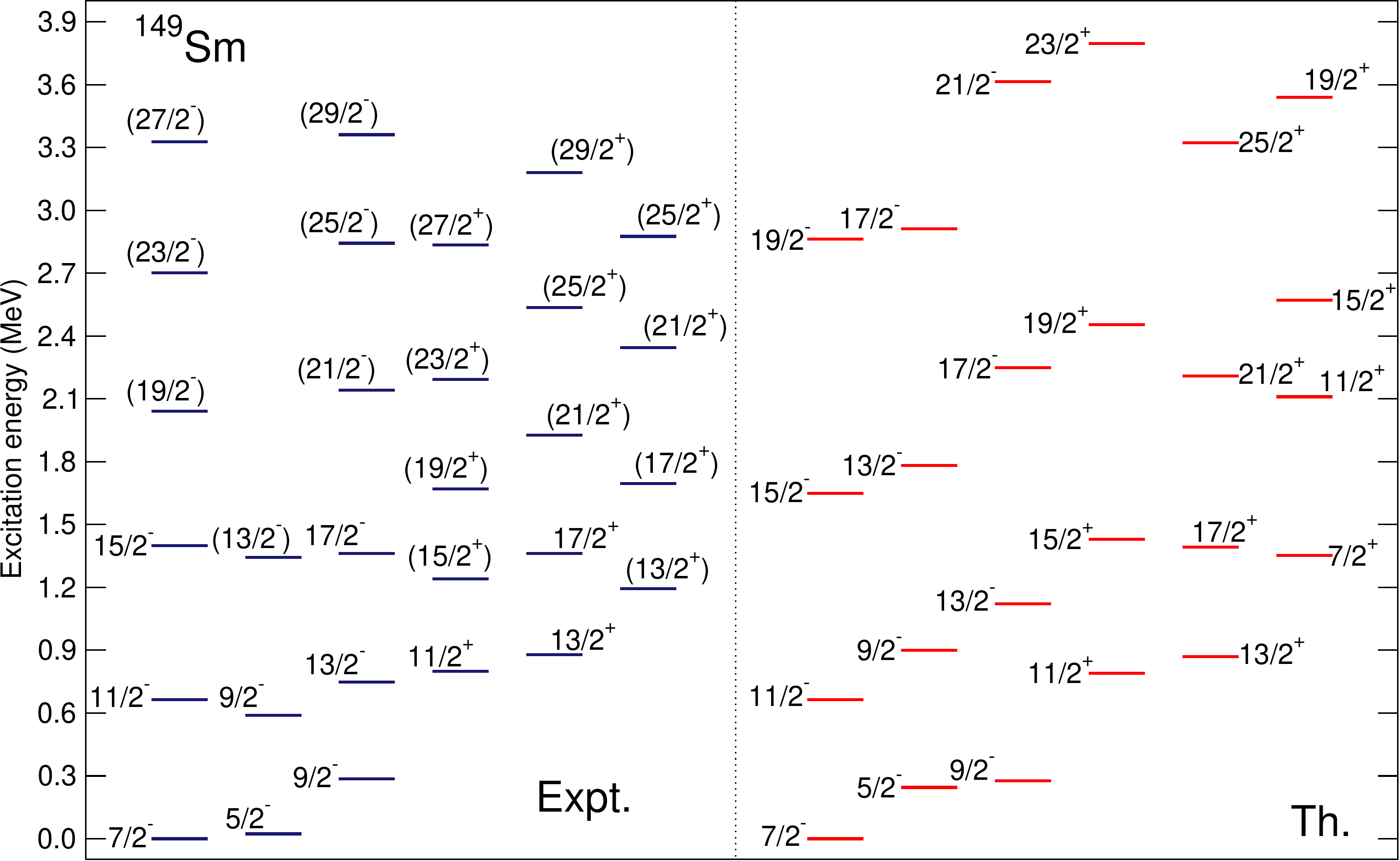} 
\caption{(Color online) The lowest three negative- and positive-parity bands of $^{149}$Sm. 
Available data are compared to model calculation in which states are classified in bands 
according to dominant E2 transitions. The data are 
 from Ref.~\cite{data}, and parentheses denote states with only a 
 tentative assignment of spin and parity.}
\label{fig:149sm}
\end{center}
\end{figure*}

\begin{figure*}[htb!]
\begin{center}
 \includegraphics[width=0.7\linewidth]{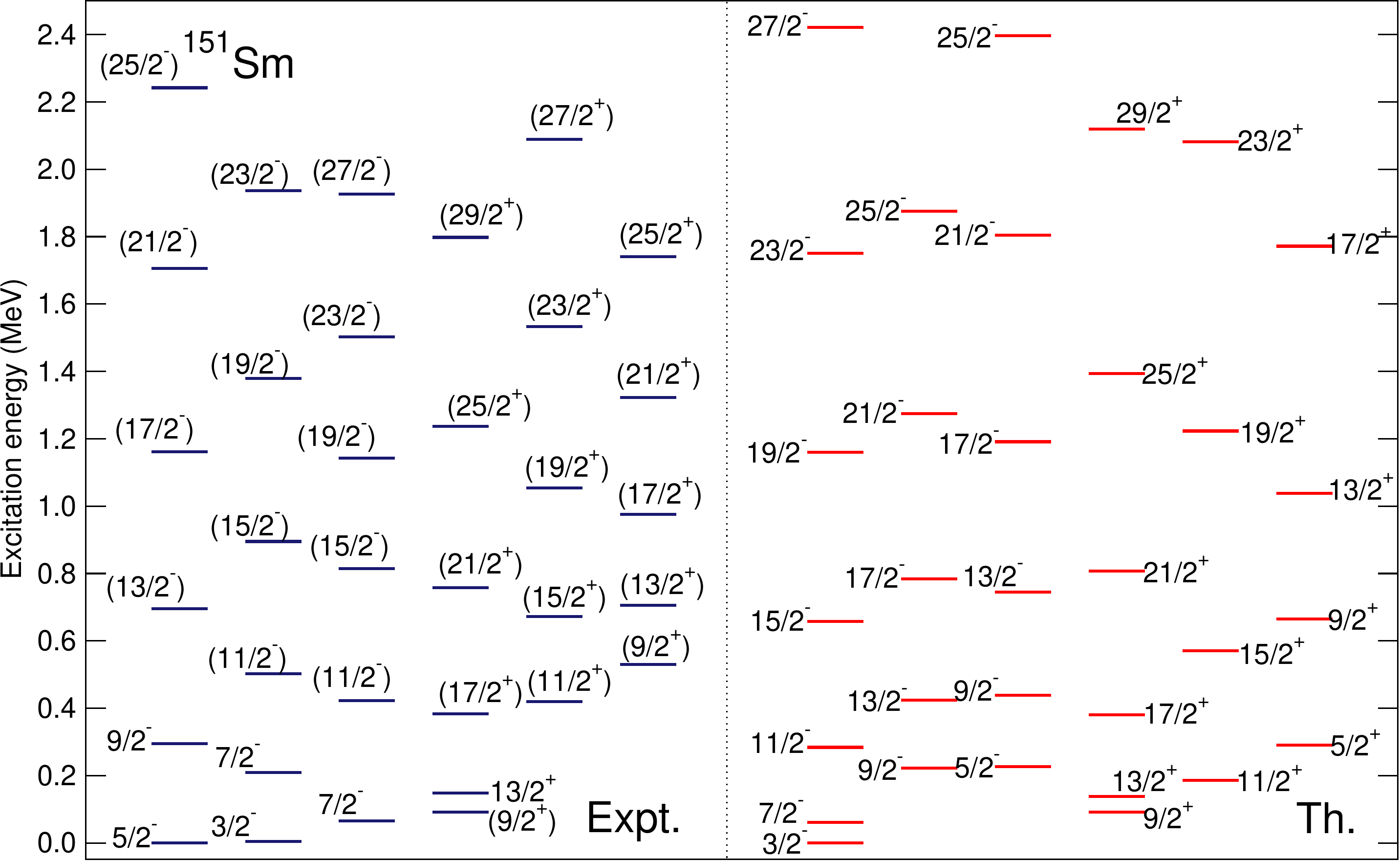} 
\caption{(Color online) Same as in the caption to Fig.~\ref{fig:149sm}, but for
 $^{151}$Sm. }
\label{fig:151sm}
\end{center}
\end{figure*}

\begin{figure*}[htb!]
\begin{center}
 \includegraphics[width=0.7\linewidth]{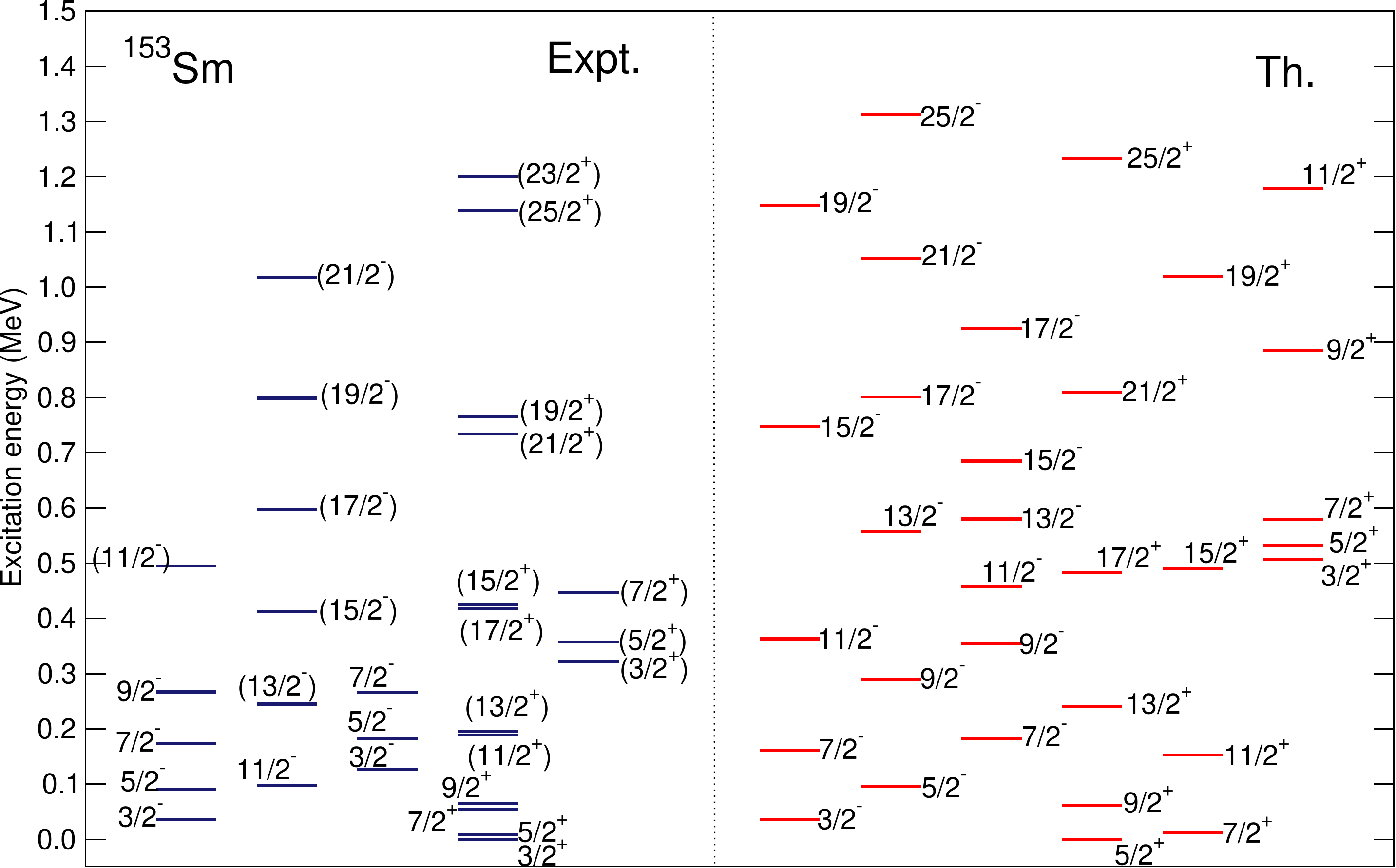} 
\caption{(Color online) Same as in the caption to Fig.~\ref{fig:149sm}, but for
 $^{153}$Sm. }
\label{fig:153sm}
\end{center}
\end{figure*}

\begin{figure*}[htb!]
\begin{center}
 \includegraphics[width=0.7\linewidth]{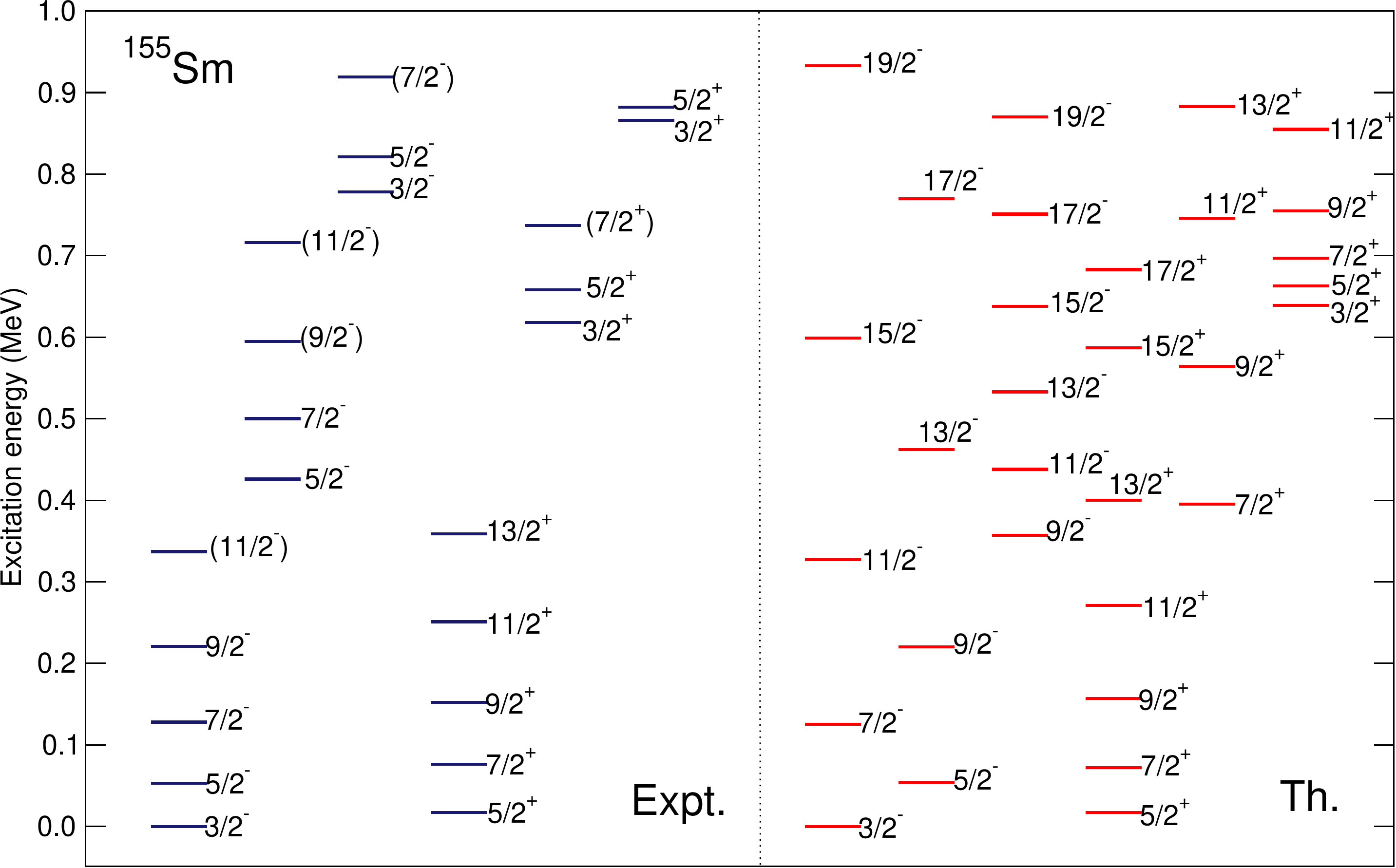} 
\caption{(Color online) Same as in the caption to Fig.~\ref{fig:149sm}, but for
 $^{155}$Sm. }
\label{fig:155sm}
\end{center}
\end{figure*}

\begin{figure*}[htb!]
\begin{center}
 \includegraphics[width=0.8\linewidth]{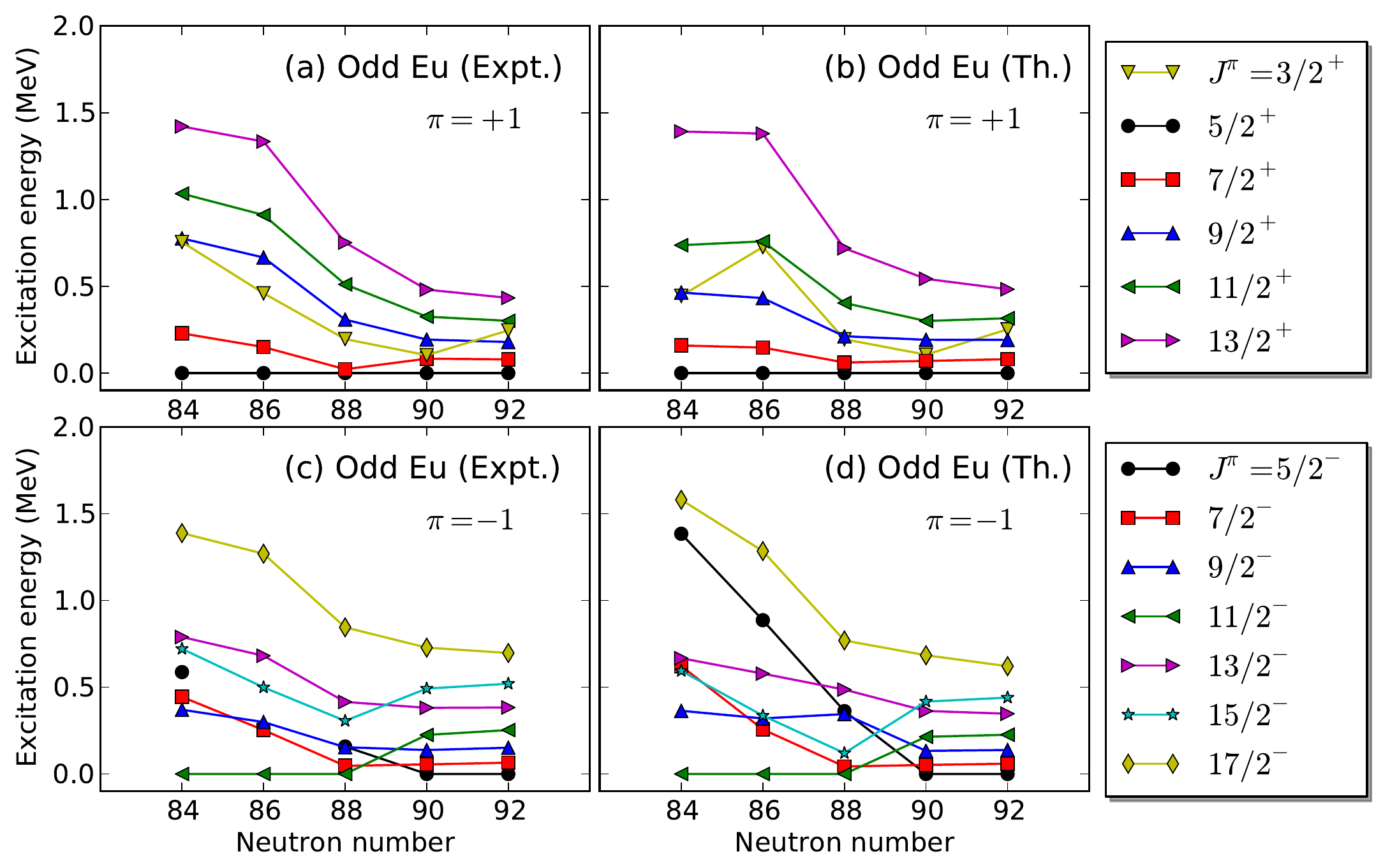} 
\caption{(Color online) Evolution of excitation energies of low-lying (a,b) positive- ($\pi=+1$) and (c,d) negative-parity ($\pi=-1$) yrast states as functions of neutron number in the isotopes $^{147-155}$Eu, in comparison with available data taken from Ref.~\cite{data}.}
\label{fig:eu-level}
\end{center}
\end{figure*}

\begin{figure*}[htb!]
\begin{center}
 \includegraphics[width=0.8\linewidth]{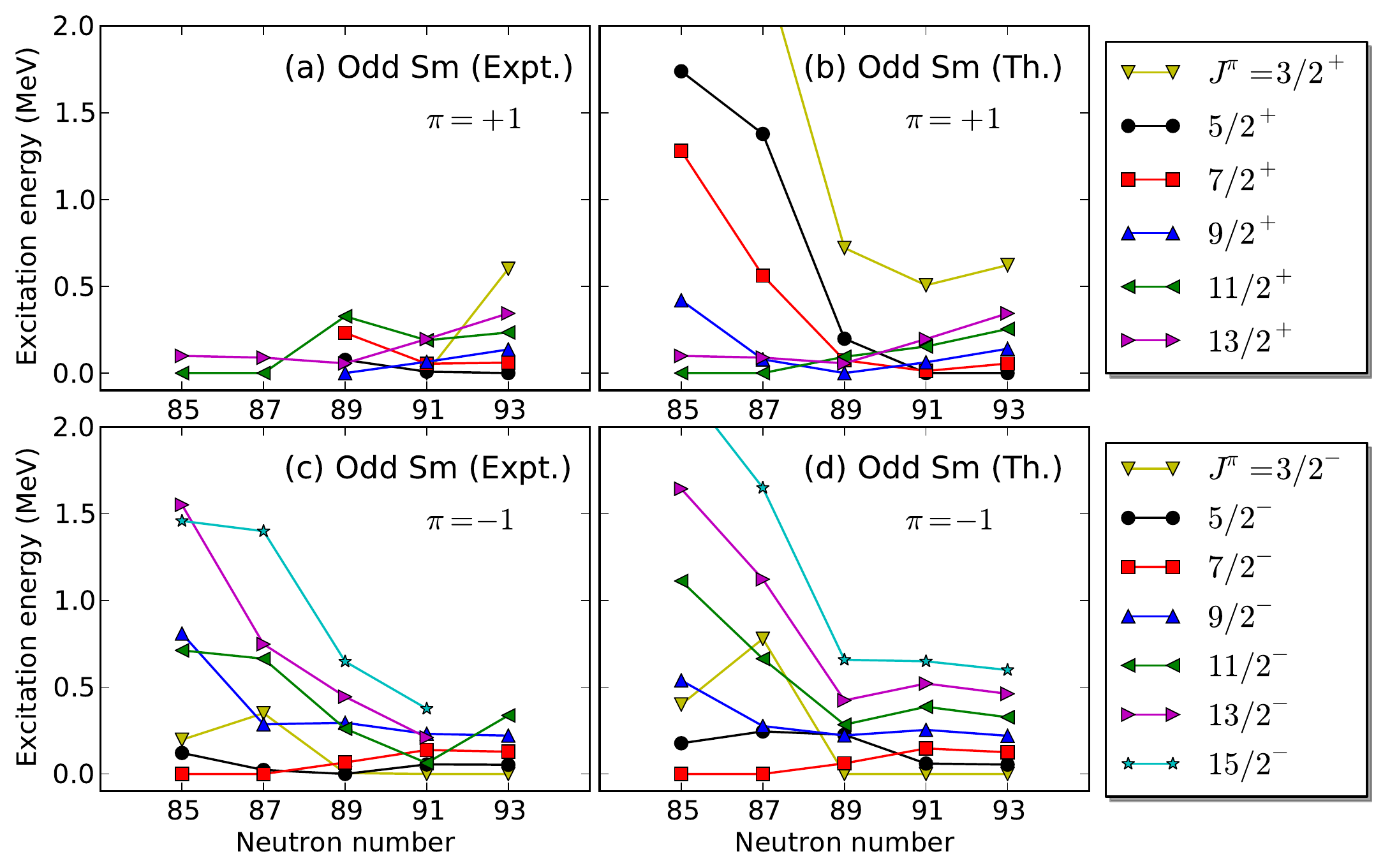} 
\caption{(Color online) Same as in the caption to Fig.~\ref{fig:eu-level}, but for the isotopes $^{147-155}$Sm. }
\label{fig:sm-level}
\end{center}
\end{figure*}

A phase transition is characterised by a significant variation of one or more order
parameters as functions of the control parameter. Even though shape phase transitions in nuclei 
have been explored extensively by considering potential energy surfaces, the deformation parameters that 
characterise these surfaces are not observables and, therefore, a quantitative study 
of QPT must go beyond the simple Landau approach and include the direct computation of 
observables related to order parameters. In the following we will consider 
spectroscopic properties of odd-mass Eu and Sm isotopes that can be associated with order parameters of 
a shape phase transition. 
To illustrate the capability of the present method to
describe low-energy spectra of odd-mass nuclei in this region, 
in Figs.~\ref{fig:149sm}-\ref{fig:155sm} we show results for the three lowest 
positive- and negative-parity bands of $^{149,151,153,155}$Sm isotopes, 
in comparison with the available experimental values \cite{data}. 
The calculated energy levels are grouped into bands according to the dominant
E2 decay pattern. We note that the energy spectrum of $^{147}$Sm is
very similar to that of the adjacent nucleus $^{149}$Sm, and available 
data are not sufficient for a detailed comparison. The calculated excitation spectra of odd-mass 
Eu nuclei have already been compared to data in Ref.~\cite{nomura2016odd}, 
including E2 and M1 transition rates, spectroscopic quadrupole moments and magnetic moments, and this is why 
these spectra are not explicitly shown here. 

For $^{149}$Sm, in Fig.~\ref{fig:149sm} one notices  that  both
positive- and negative-parity bands exhibit a vibrational level structure 
characterised by the $\Delta J=2$ systematics of the weak-coupling
limit. The agreement between the calculated and experimental spectra is
fairly good. The excitation spectrum of $^{151}$Sm, shown in Fig.~\ref{fig:151sm}, 
is less harmonic and the levels are more compressed in energy. 
All bands, however, still display the $\Delta J=2$ structure 
indicating that the odd neutron is not strongly coupled to the boson core.

A significant change in the structure of excitation spectra occurs between $^{151}$Sm and $^{153}$Sm 
(Fig.~\ref{fig:153sm}). Both in experiment and model calculations $\Delta J=1$ bands characteristic 
for the strong-coupling limit are formed, and coexist with 
bands that exhibit the weak-coupling $\Delta J=2$ systematics. 
In addition, $^{153}$Sm is the only nucleus among all the odd-mass Sm
isotopes considered in which the ground state has positive
parity, originating from the neutron $1i_{13/2}$ orbit. The spectrum of $^{153}$Sm 
reflects the abrupt change of structure at $N = 90$ in the even-even boson core,  
that can approximately be characterised by the X(5) analytic solution at the critical point of the 
first-order quantum phase transition between spherical and axially deformed shapes \cite{li2009}.
We note that in experiment all the negative-parity bands display the 
strong-coupling structure, whereas both $\Delta J=2$ and $\Delta J=1$ sequences 
of negative-parity states are obtained in the model calculation. This discrepancy can  
most probably be attributed to the occupation probabilities of the corresponding 
single-particle orbitals obtained in the SCMF calculation. 
In addition, some structures could be based on intruder orbitals that develop from the shell below
the neutron $N=82$ closure and, therefore, beyond the model space
in which the IBFM Hamiltonian is diagonalised. For instance, the experimental band 
built on the state ${11/2}^-_1$ has been attributed to the
neutron configuration ${11/2}^-[505]$ \cite{data}. 

The transition to a well deformed, axially symmetric shape, is completed in 
$^{154}$Sm (cf. Fig.~\ref{fig:pes}), and this is clearly reflected in the excitation spectrum of 
$^{155}$Sm, shown in Fig.~\ref{fig:155sm}. Sequences of both negative- and positive-parity 
states exhibit the $\Delta J=1$ structure of the
strong-coupling limit and the excitation energies follow to a good approximation 
the simple rotational $J(J+1)$ pattern. 

We note that, as already shown in our previous article of Ref.~\cite{nomura2016odd}, a similar level
of quantitative agreement with data is obtained for the odd-mass Eu isotopes. 
The band structure of Eu nuclei is simpler than that of
the corresponding odd Sm isotopes: in $^{147-151}$Eu the lowest three positive-parity
bands follow the $\Delta J=1$ systematics of the strong-coupling limit,
while the lowest three negative-parity bands exhibit the weak-coupling
$\Delta J=2$ structure; in $^{153,155}$Eu 
the lowest three bands of both positive- and negative-parity are characterised by the
strong-coupling $\Delta J=1$ sequence of states.

To analyze the overall systematics of excitation spectra in the transition from spherical to deformed equilibrium shapes, in Figs.~\ref{fig:eu-level} and \ref{fig:sm-level} we plot the calculated spectra for the low-lying positive ($\pi=+1$) and negative parity ($\pi=-1$) yrast states in the isotopes $^{147-155}$Eu and $^{147-155}$Sm, as functions of neutron number, and compare them with available data \cite{data}. 
For both odd-$A$ Eu and Sm isotopes the model reproduces the experimental systematics, except for a few states in odd-$A$ Sm isotopes with $N=89$ or 91. 
The phase transition is characterized by a change in the spin of the ground state for a particular nucleus \cite{petrellis2011}. 
Indeed one notices that the ground-state spin changes at $N=90$ in the Eu isotopes for negative parity, and at $N=89$ in odd-$A$ Sm for both parities. For the positive-parity states in Eu, however, the change does not occur and the ${5/2}^+$ level remains the ground state for all isotopes.


\begin{figure*}[htb!]
\begin{center}
\includegraphics[width=\linewidth]{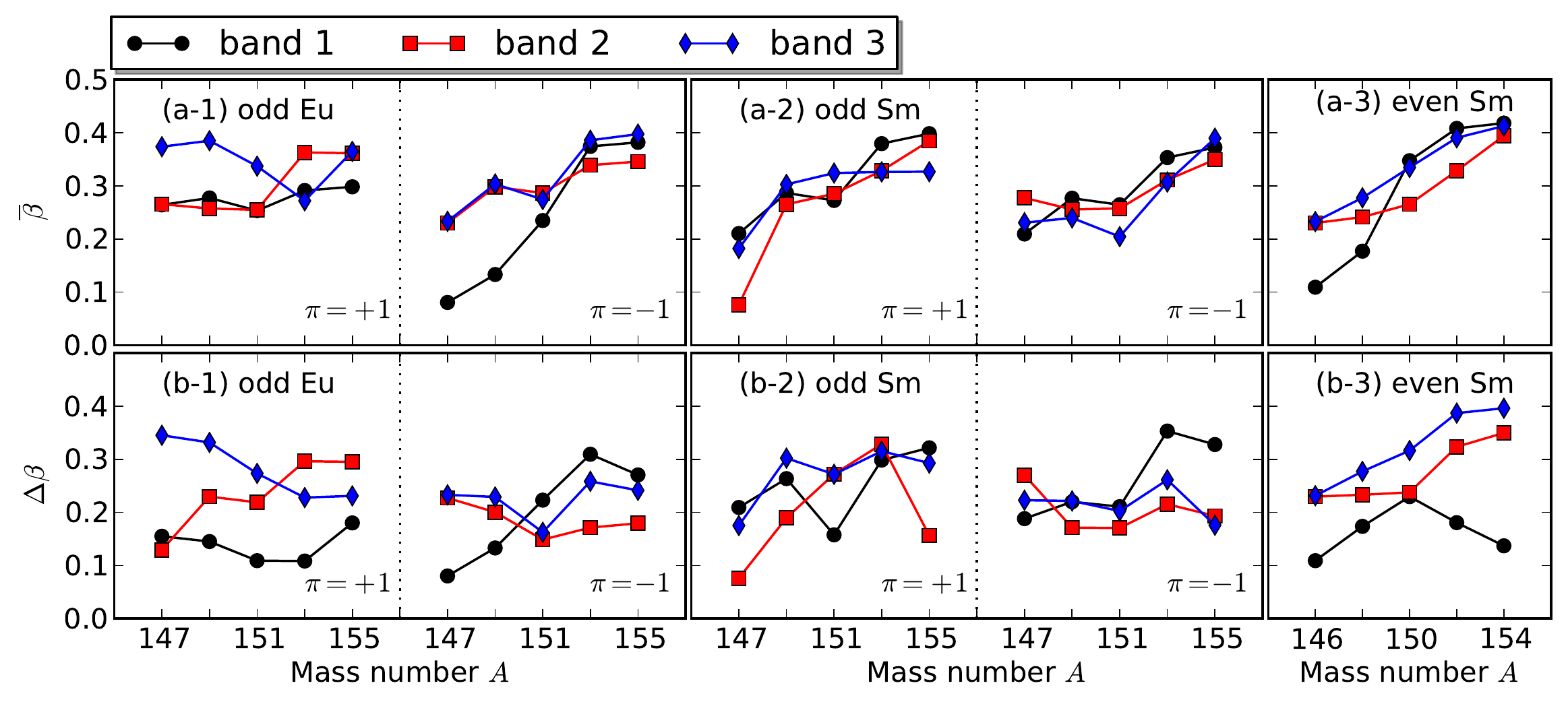} 
\caption{(Color online) Evolution of characteristic mean-field quantities calculated
 for the lowest three positive- ($\pi=+1$) and negative-parity 
 ($\pi=-1$) bands of the odd-$A$ Eu and Sm nuclei, and  $K^{\pi}=0^+$ bands in the the even-A Sm isotopes, as 
 functions of mass number: the mean value $\overline{\beta}$ (a) and
 variance $\Delta\beta$ (b) of the equilibrium deformation
 parameter for the bandhead state. ``band 1'', ``band 2'' and ``band 3''
 denote the lowest, second- and third-lowest bands, respectively. See the
 main text for the definition of each quantity. } 
\label{fig:def}
\end{center}
\end{figure*}

In the remainder of this section possible signatures of QPTs are explored 
in odd-mass Eu (odd proton) and Sm (odd neutron) nuclei at $N \approx 90$.  
We start by considering the equilibrium axial deformation parameter $\beta$ which, even though  
it is not an observable, can nevertheless be used in a theoretical analysis to describe the 
evolution of deformation with the control parameter and as a 
signature of QPT in both even-even and odd-mass systems, as shown in the classical
study of Ref.~\cite{iachello2011} using the IBFM. However, 
in contrast to a mean-field description of QPT based on the
analysis of potential energy surfaces around equilibrium minima, 
we explicitly compute the deformation parameter
for a given state using the wave function obtained by diagonalising the IBFM Hamiltonian $\hat H$ in Eq.~(\ref{eq:ham}). 
Figure \ref{fig:def} displays the mean value of axial quadrupole
deformation $\overline{\beta}=\sqrt{\langle\beta^2\rangle}$ (a-1,a-2,a-3) and the 
variance $\Delta\beta=\sqrt{\langle\beta^2\rangle-\langle\beta\rangle^2}$
(b-1,b-2,b-3) for the considered odd Eu and Sm isotopes, as well as the corresponding even-even Sm cores,
calculated for the band-heads of three lowest positive ($\pi=+1$) and
negative-parity ($\pi=-1$) bands. 
Note that for the even-even Sm nuclei all values included in Fig.~\ref{fig:def}
correspond to the lowest three $K^{\pi}=0^+$ bands, where 
$K^{\pi}$ denotes the projection of the total angular
momentum on the symmetry axis of the intrinsic frame. 
The expectation value $\langle\beta^{\lambda}\rangle$ ($\lambda=1,2$)
for the state with spin and parity $J^{\pi}$, which is used 
in the calculation of $\overline{\beta}$ and $\Delta\beta$, is defined
by the relation: 
\begin{eqnarray}
\label{eq:av}
 \langle\beta^{\lambda}\rangle
=\langle\Psi_{J,k}|\beta^{\lambda}|\Psi_{J,k}\rangle
=\int
d\beta\langle\Psi_{J,k}|\phi_{J,k}(\beta)\rangle\beta^{\lambda}\langle\phi_{J,k}(\beta)|\Psi_{J,k}\rangle
=\int d\beta\beta^{\lambda}|\langle\phi_{J,k}(\beta)|\Psi_{J,k}\rangle|^2
; \,
\end{eqnarray}
where $|\Psi_{J,k}\rangle$ is the eigenstate of the IBFM Hamiltonian,
with $k$ distinguishing states with the same 
$J$, and $|\phi_{J,k}(\beta)\rangle$ is the projected intrinsic state 
of the coupled boson-fermion system \cite{paar1982,leviatan1988}. 
In the integral of Eq.~(\ref{eq:av}) the expectation value is computed in the interval $|\beta|<0.6$. 
Values of $\beta$ larger than $0.6$ are not relevant because of the restricted boson 
model space built from a limited number of valence nucleon pairs. 
Consistent with the evolution of the equilibrium minimum at the
mean-field level (cf. Fig.~\ref{fig:pes}), the average deformation $\overline{\beta}$ in most
cases increases monotonically with nucleon number to a value
of approximately 0.35 for heavier isotopes. In the odd-mass
Eu and Sm nuclei one notices a significant change from $A=151$ to 153. 
Similarly, the variance $\Delta\beta$ changes (either increases or
decreases) mostly for the transitional nuclei with $A=151$ or 153.  
The calculated values of $\overline{\beta}$ and $\Delta\beta$ evolve as 
expected, that is, the fluctuations in shape variables increase as a  
result of comparatively softer potentials in transitional nuclei. 

\begin{figure*}[htb!]
\begin{center}
\includegraphics[width=0.7\linewidth]{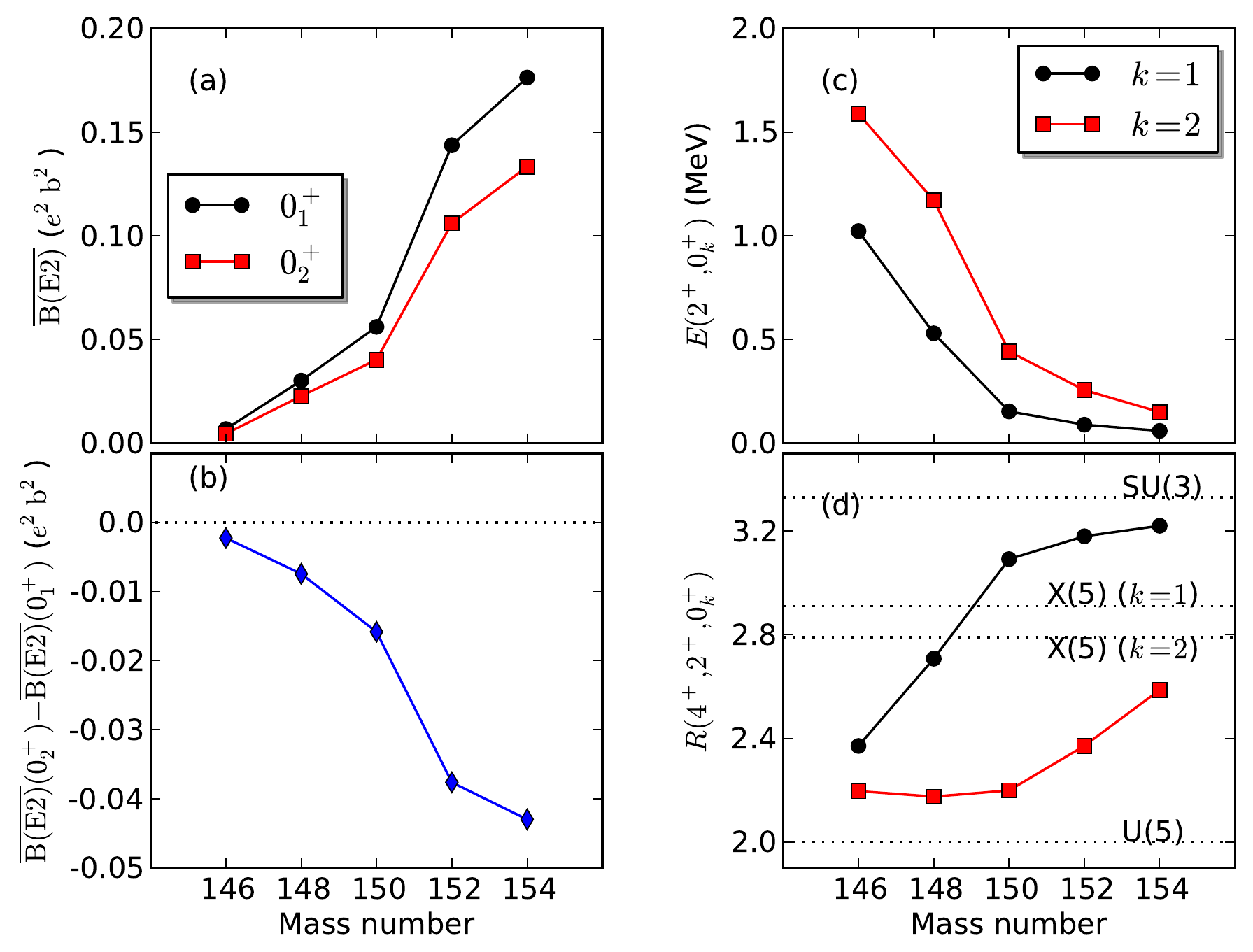} 
\caption{(Color online) Left: calculated values of $\overline{\rm
 B(E2)}$ for the two lowest $0^+$ states (a), and the 
 difference $\overline{\rm B(E2)}(0^+_2)-\overline{\rm B(E2)}(0^+_1)$
 (b) for the even-even Sm isotopes. Right: the excitation energies $E(2^+,0^+_k)$ ($k=1,2$) of the states $2^+$ (c) and
 the energy ratio of the $4^+$ to the $2^+$ excited states $R(4^+,2^+,0^+_k)$ (d) for the
 $K^{\pi}=0^+_1$ and $0^+_2$ bands of the even-even core nuclei $^{146-154}$Sm. In
 panel (d) the corresponding 
 values of the energy ratio $E(4^+_1)/E(2^+_1)$ in the
 U(5), X(5) and SU(3) limits: 2.00, 2.91 ($k=1$), 2.79 ($k=2$)
 and 3.33, respectively, are also indicated by dotted lines. }
\label{fig:qinvariant-even}
\end{center}
\end{figure*}

\begin{figure*}[htb!]
\begin{center}
\includegraphics[width=\linewidth]{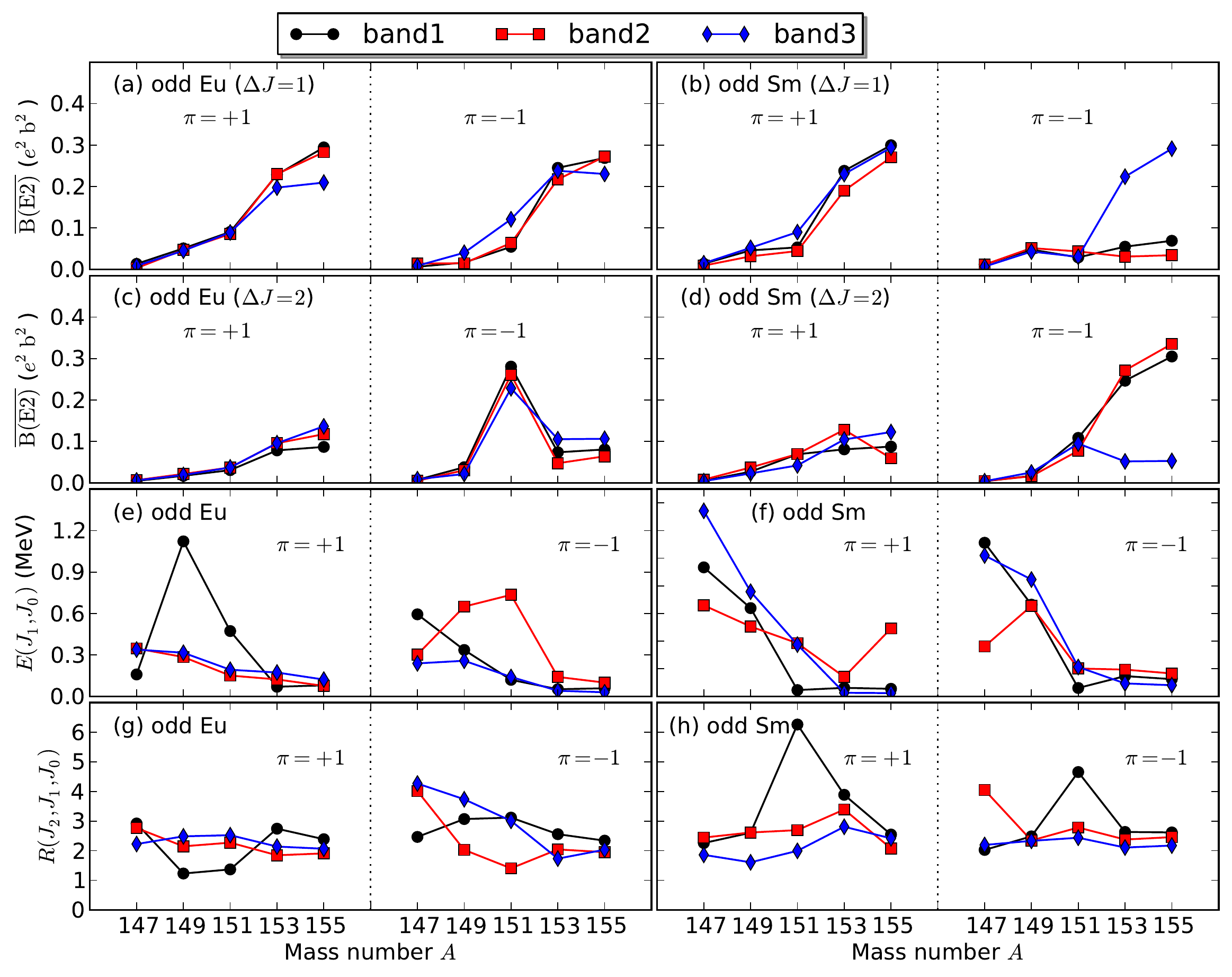} 
\caption{(Color online) Dependence on mass number of the
 calculated spectroscopic quantities for the odd-$A$ Eu and Sm isotopes: 
average $B({\rm E2}; J_0+\Delta J \rightarrow J_0)$ 
 (in units of $e^2{\rm b}^2$) for transitions 
  between the band-head $J_0$ of a given band and the 
 lowest five states with $J_0+\Delta J$, with $\Delta J=1$ (a,b) and 2 (c,d); the excitation 
 energy $E(J_1,J_0)$ ($J_1=J_0+\Delta J$, in MeV) of the second-lowest state in a band relative 
 to the band-head (e,f), and the energy ratio $R(J_2,J_1,J_0)$ with
 $J_2=J_0+2\Delta J$ (g,h). 
See the main text for the definition of each quantity. }
\label{fig:qinvariant}
\end{center}
\end{figure*}

The evolution of E2 transition rates with neutron number can also indicate a sudden 
change of deformation. In analogy to the quadrupole shape invariant $q_2$ which provides a measure 
of axial deformation in even-even nuclei \cite{kumar1972,cline1986}, here we
consider the quantity  $\overline{\rm B(E2)}$, 
defined as the average $B$(E2) for transitions between the band-head of a given band
with spin $J_0$ and the lowest $n$ states with 
spin $J_0+\Delta J$: 
\begin{eqnarray}
\label{eq:qinvariant}
 \overline{\rm B(E2)}=\frac{1}{n}\sum_{k=1}^n B({\rm E2}; (J_0+\Delta J)_k
  \rightarrow J_0), 
\end{eqnarray}
where $\Delta J=1$ or 2, and the sum is in order of increasing excitation energies 
of the levels $J_0+\Delta J$. Only a few lowest transitions 
will contribute significantly to this quantity and, therefore, 
$n=5$ terms have been included in the sum. 
In the following the average $B$(E2) transition defined in 
Eq.~(\ref{eq:qinvariant}) is referred to as q-invariant. 

In the case of even-even nuclei
quadrupole shape invariants have been used to
quantify the first-order QPT between spherical and axially deformed
shapes. In Ref.~\cite{werner2008}, based on the available data on deformed Gd
isotopes and on a schematic IBM-1 calculation, the crossing of 
q-invariants of the ground and
first excited $0^+$ states has been shown to occur near the point of
shape phase transition.
In Fig.~\ref{fig:qinvariant-even}(a,b) we plot 
the evolution of $\overline{\rm B(E2)}$ for the ground ($0^+_1$) and
first excited $0^+_2$ states in the even-even Sm isotopes, and their difference as functions of mass number.
Even though the present model calculation does
not exhibit the crossing of the q-invariants for $0^+_1$ and $0^+_2$, 
some basic features are
still observed in Fig.~\ref{fig:qinvariant-even}(a,b): for the lighter isotopes the two q-invariants are almost 
identical in magnitude and display a similar increase with mass number whereas, starting from the point of phase 
transitions between $^{150}$Sm and $^{150}$Sm, their difference $\overline{\rm B(E2)}(0^+_2)- \overline{\rm B(E2)}(0^+_1)$ 
increases rapidly in magnitude and the q-invariant 
(deformation) of the ground state becomes considerably larger. 

To analyse how the odd particle influences the location and nature of the
QPT observed in the even-even core, in Fig.~\ref{fig:qinvariant}(a-d) we display the
calculated values of the q-invariant $\overline{\rm B(E2)}$ of bandhead
states for the three lowest positive- ($\pi=+1$) and negative-parity
($\pi=-1$) bands in the odd-$A$ Eu and Sm isotopes, for the two cases of
$\Delta J=1$ and $\Delta J=2$ transitions. 
In analogy to the even-even case shown in
Fig.~\ref{fig:qinvariant-even}(a,b), one expects the
q-invariants to increase with neutron number. 
This feature does not seem to generally apply in the present calculation on the odd-$A$
Eu and Sm nuclei, rather it depends on whether a band exhibits a 
weak-coupling or strong-coupling systematics. 
In the negative-parity bands of the odd-$A$ Sm nuclei in
Fig.~\ref{fig:qinvariant}(b,d), for instance, the q-invariants of the
band-head states of the lowest and 
second-lowest bands do not display a notable change from A=151 to 155
in the $\Delta J=1$ case, whereas the q-invariant of the band-head
state of the third-lowest band increases. 
In the $\Delta J=2$ case the opposite is observed: the q-invariants
of the lowest and second-lowest bands increase with neutron number, whereas that
of the third-lowest band remains almost unchanged. 
As shown in Figs.~\ref{fig:153sm} and \ref{fig:155sm}, the lowest two
negative-parity bands of $^{153,155}$Sm follow the 
$\Delta J=2$ decay systematics typical of the weak-coupling limit, whereas the 
third-lowest negative-parity band follows the strong-coupling 
systematics and E2 transitions with $\Delta J=1$ dominate. 
Nevertheless, one expects that the most significant change (either
an increase or decrease) 
of the q-invariants occurs between A=149 and 151, or between A=151
and 153, in accordance with the even-even case exhibiting an abrupt 
change of this quantity between A=150 and 152. 

Shape transitions can be also characterised by the evolution of excitation energies, as shown in 
Fig.~\ref{fig:qinvariant}(e,f) where we plot, for the three lowest bands of both parities, the energy
difference
\begin{eqnarray}
 E(J_1,J_0) = E(J_1) - E(J_0).
\end{eqnarray}
$E(J_0)$ and $E(J_1)$ ($J_1=J_0+\Delta J$ with $\Delta J=1,2$) are the energies of the
band-head and the first excited state in a band, respectively. 
With the exception of an increase from mass $A=147$ to 149 for two bands in odd-$A$ Eu 
isotopes that can probably be attributed to a more pronounced band mixing, 
the quantity $E(J_1,J_0)$ decreases with neutron number, with a rapid change 
in the transitional nuclei at $N\approx 90$. 
Indeed, the $2^+$ excitation energies in the bands $K^{\pi}=0^+_1$ and $0^+_2$ belonging to the 
corresponding even-even Sm nuclei and plotted in
Fig.\ref{fig:qinvariant-even}(c), exhibit a sudden decrease from A=148
to 150 that reflects the abrupt rise of deformation. 
As another signature of the shape phase transition related to excitation energies, 
we consider the energy ratio between the lowest two excited
states (with spin $J_1=J_0+\Delta J$ and $J_2=J_0+2\Delta J$) in a given band: 
\begin{eqnarray}
R(J_2,J_1,J_0)=\frac{E(J_2)-E(J_0)}{E(J_1)-E(J_0)}. 
\end{eqnarray}
In the even-even case this is nothing but the
ratio of the $4^+$ to $2^+$ excitation energies in $K^{\pi}=0^+$ bands and,
especially the ratio in the yrast band,
$R(4^+_1,2^+_1,0^+_1)=E(4^+_1)/E(2^+_1)$ has often been used as a signature
of phase transition between vibrational and rotational nuclei. 
In Fig.~\ref{fig:qinvariant-even}(d) we plot the evolution of the ratio
$R(4^+,2^+,0^+)$ for the lowest two bands with $K^{\pi}=0^+$ in the even-A Sm 
isotopes as function of the mass (neutron) number. 
The ratio $R(4^+_1,2^+_1,0^+_1)$ exhibits a typical increase as a function of A, from
close to the vibrational (or U(5)) limit ($R(4^+_1,2^+_1,0^+_1)=2.00$), to the rotational
(or SU(3)) limit ($R(4^+_1,2^+_1,0^+_1)=3.33$). 
The $R(4^+_1,2^+_1,0^+_1)$ value of 2.91, predicted by the X(5) critical-point
symmetry model for the phase transition \cite{X5}, is located between A=148 and 150. 
The ratio $R(4^+,2^+,0^+_2)$ also exhibits an increase, but is
always smaller than $R(4^+_1,2^+_1,0^+_1)$, and smaller than the value predicted by the 
X(5) model. The value of the ratio $R(4^+,2^+,0^+_2)$ differs from 
that of $R(4^+_1,2^+_1,0^+_1)$ particularly in the transitional nuclei, reflecting the 
different intrinsic structures of the two lowest $0^+$ states. 
A similar trend is observed in the odd-$A$ nuclei: 
Figs.~\ref{fig:qinvariant}(g) and \ref{fig:qinvariant}(h) show the calculated
values of the ratio $R(J_2,J_1,J_0)$ for the lowest three positive- and negative-parity
bands in the considered odd-$A$ Eu and Sm isotopes, respectively. 
In the vibrational (A=147 and 149) and deformed rotational (A=153 and
155) nuclei, the calculated ratios exhibit similar values in all bands of a given parity, 
but differ significantly in the transitional nuclei with A=151. 

\begin{figure*}[htb!]
\begin{center}
\includegraphics[width=\linewidth]{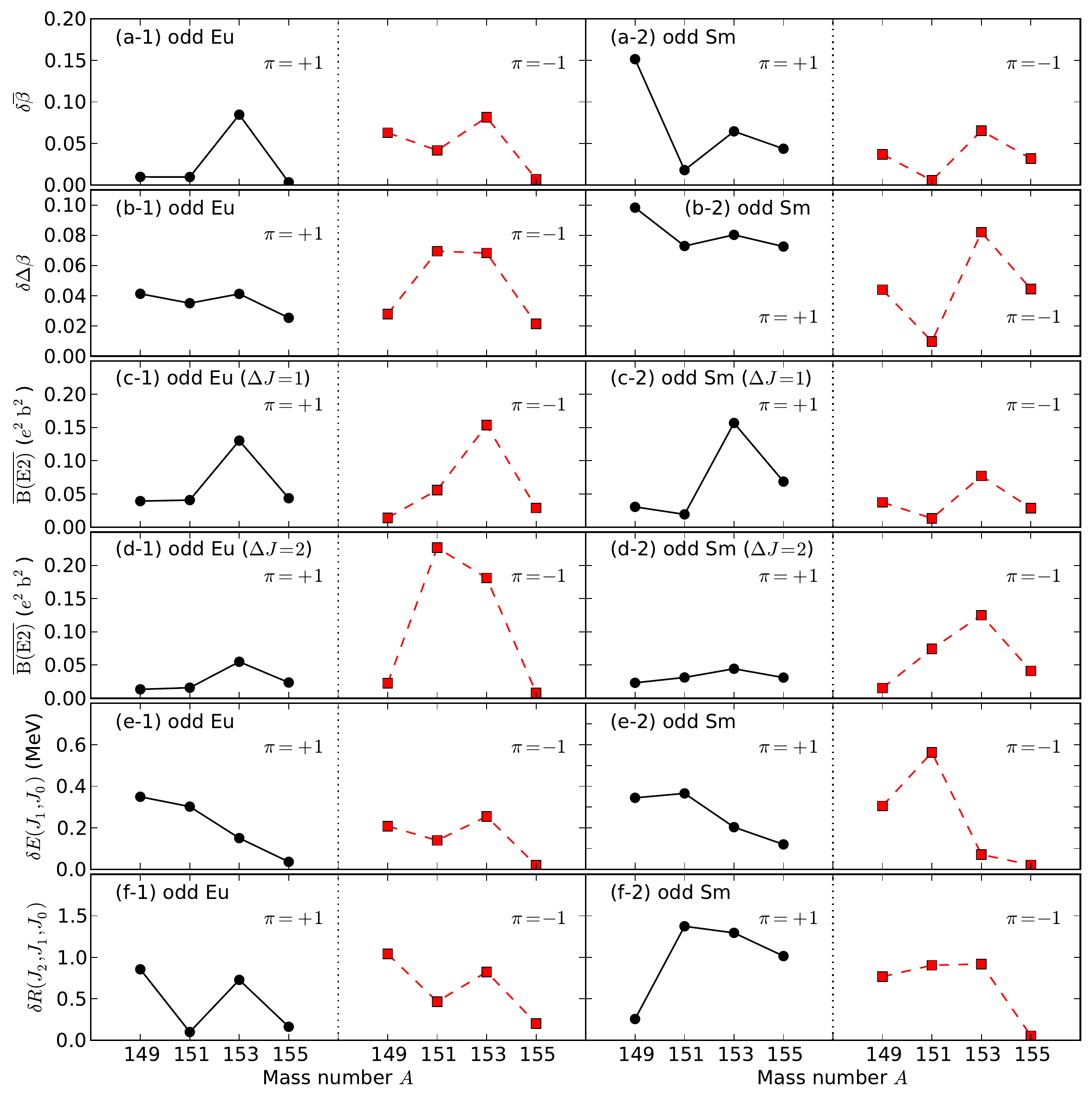} 
\caption{(Color online) Differentials of the mean value
 $\delta\overline\beta$ (a-1, a-2) and variance $\delta\Delta\beta$
 (b-1, b-2) of the quadrupole deformation parameter $\beta$,
 q-invariants $\overline{\rm B(E2)}$ with $\Delta J=1$ (c-1, c-2) and
 $\Delta J=2$ (d-1, d-2), excitation energy $\delta E(J_1,J_0)$ (e-1, e-2)
 and the energy ratio $\delta R(J_2,J_1,J_0)$, for the odd-$A$ Eu
 and Sm isotopes, as functions of the mass number. }
\label{fig:diff}
\end{center}
\end{figure*}

\begin{figure*}[htb!]
\begin{center}
\includegraphics[width=8.6cm]{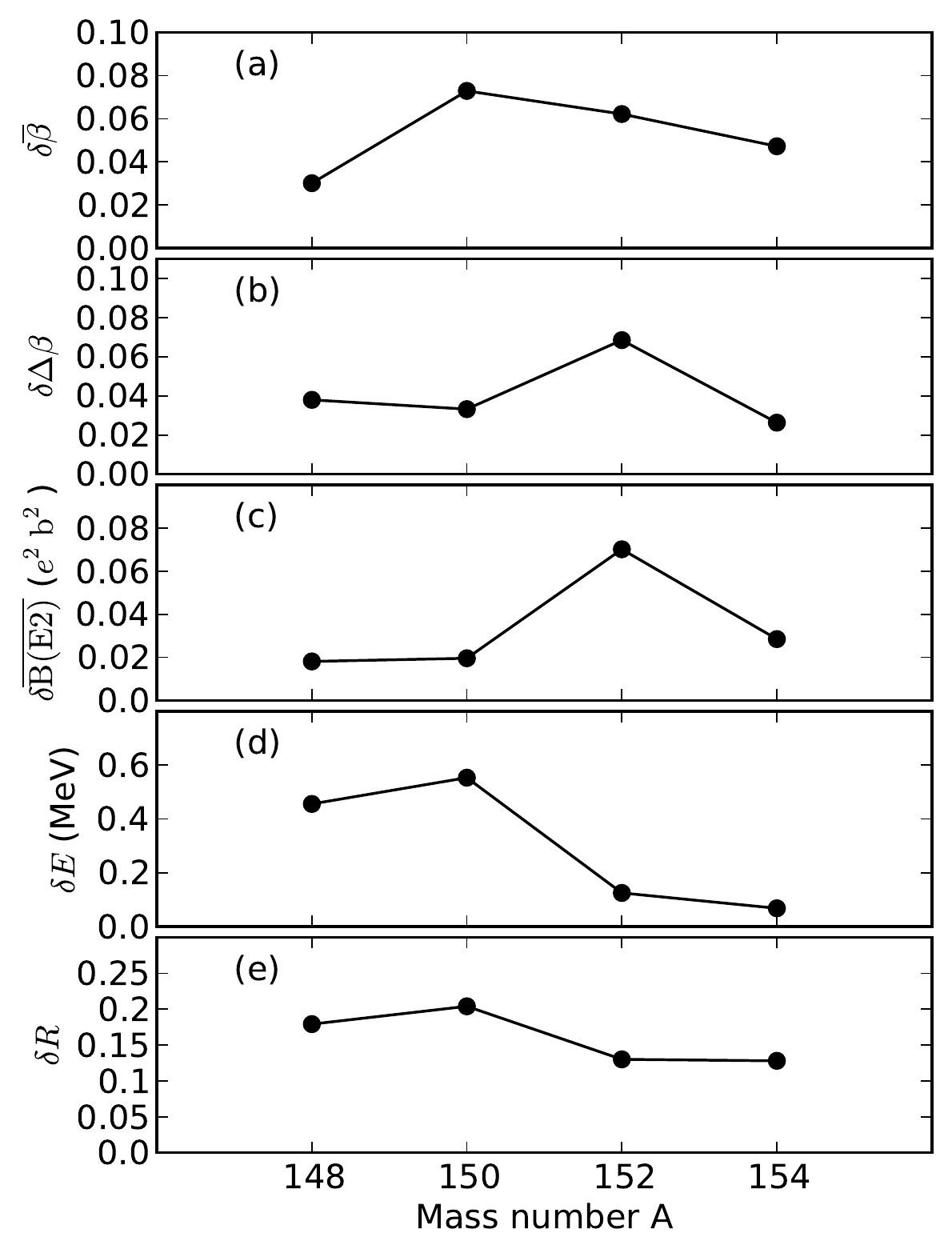} 
\caption{Same as in the caption to Fig.~\ref{fig:diff}, but for the
 even-even Sm core nuclei. }
\label{fig:diff-even}
\end{center}
\end{figure*}

We have shown that the characteristic mean-field and spectroscopic
properties of odd-$A$ Eu and Sm nuclei as functions of the neutron number, 
as well as those of the even-even Sm isotopes,
exhibit a rapid change close to the transitional nuclei with mass A=151 or 153. 
To identify more precisely the location of discontinuities characteristic of shape
phase transitions we consider the differentials of these quantities, that
is, the difference of their values in neighbouring isotopes. 
The relevance of differential observables for studies of  
structural evolution of nuclear systems, especially in exotic nuclei,
was already pointed out in Ref.~\cite{cakirli2009}, where 
experimental differential observables related to mean-square charge
radii, spectroscopic properties, and mass observables of even-even nuclei were
analysed for different regions in the nuclear chart.  
To facilitate a similar analysis in the case of odd-$A$ nuclei in which the density of low-energy levels is much higher, 
here we define the differential of a given quantity
${\cal O}$ for a nucleus with mass $A$ as its absolute value averaged over 
the lowest bands $i$, that is, 
\begin{eqnarray}
\label{eq:diff}
 \delta{\cal O}={1\over n} \sum_{i=1}^{n}|{\cal O}_{i,A} - {\cal O}_{i,(A-2)}|
\end{eqnarray}
Figure~\ref{fig:diff} displays the differentials of the mean
value $\delta\overline{\beta}$ (a-1,a-2) and variance
$\delta\Delta\beta$ (b-1,b-2) of the quadrupole deformation $\beta$, the
q-invariants $\delta\overline{\rm B(E2)}$ in the cases of $\Delta J=1$ 
(c-1,c-2) and $\Delta J=2$ (d-1,d-2), 
the energy $\delta E(J_1,J_0)$ (e-1,e-2), and the ratio $\delta R(J_2,J_1,J_0)$
(f-1,f-2), averaged over the lowest three ($n=3$) positive- 
and negative-parity bands in the odd-$A$ Eu and Sm isotopes. 
One notices that apart 
from only a few exceptions, that is, $\delta\overline{\rm B(E2)}$ in the
case of $\Delta J=2$ for
the positive-parity states in odd Sm (Fig.~\ref{fig:diff}(d-2)) and $\delta E(J_1,J_0)$ for the 
positive-parity states in odd Eu (Fig.~\ref{fig:diff}(e-1)), the differentials of the 
considered quantities exhibit a pronounced discontinuity at the transitional nuclei,
where the potential 
becomes notably soft in both deformation parameters $\beta$ and $\gamma$ (cf. 
Fig.~\ref{fig:pes}): either at $A=151$ or 153. 
In Fig.~\ref{fig:diff-even} we plot the differentials of the same
quantities but for the even-even Sm isotopes. 
Note that the average in Eq.~(\ref{eq:diff}) is taken over the lowest three
$K^{\pi}=0^+$ bands for $\delta\overline{\beta}$ and $\delta\Delta\beta$, and
for the lowest two $K^{\pi}=0^+$ bands for $\delta\overline{\rm B(E2)}$,
$\delta E(2^+,0^+)$ and $\delta R(4^+,2^+,0^+)$. These plots clearly show 
that the differentials of the characteristic quantities in 
the even-even core nuclei also exhibit abrupt changes between the 
nuclei with mass number $A=150$ and 152, and these changes correspond to the ones 
observed in the odd-proton and odd-neutron systems.

\begin{figure}[htb!]
\begin{center}
\includegraphics[width=8.6cm]{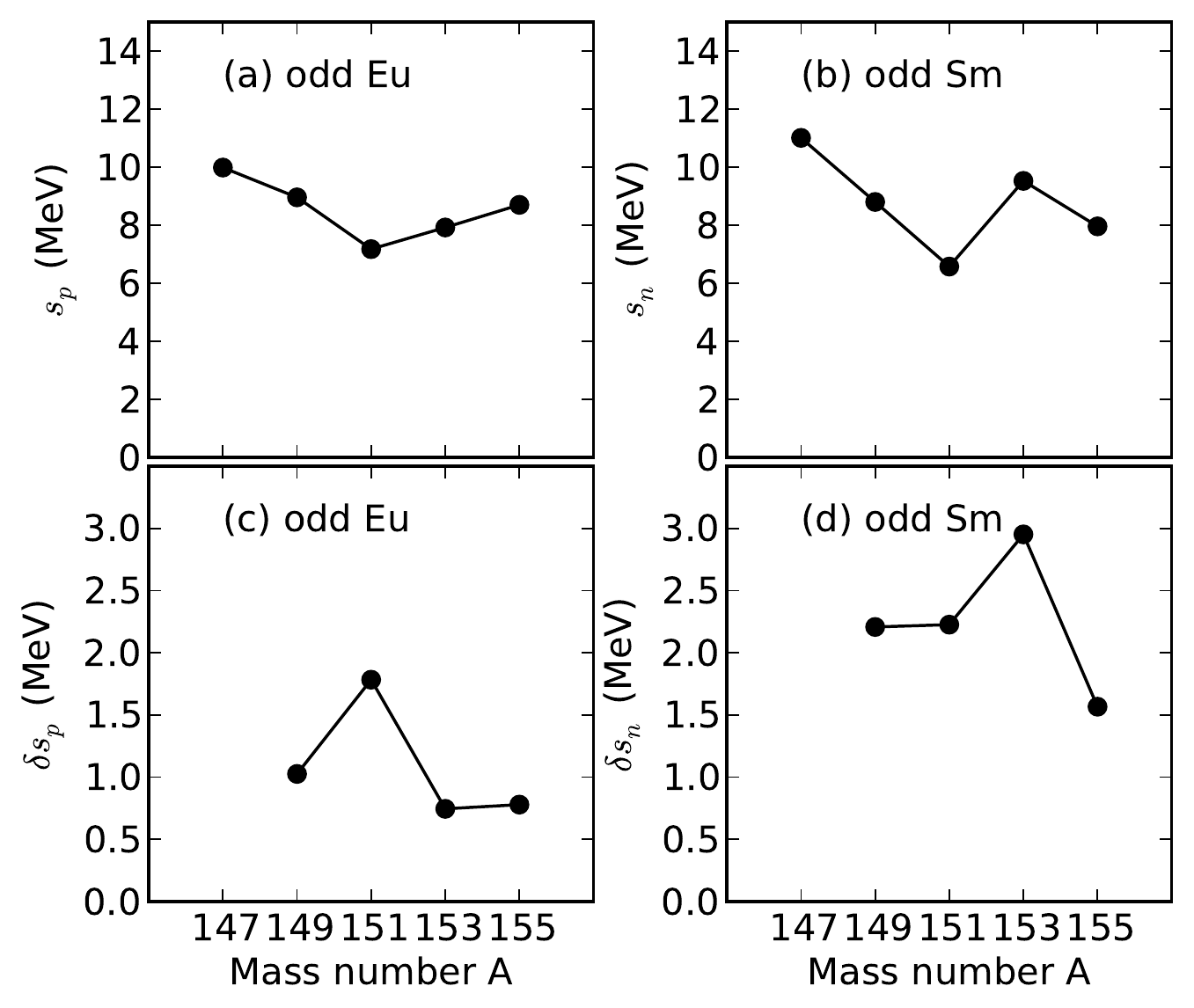} 
\caption{
 Proton (a) and neutron (b) separation energies ($s_p$ and $s_n$), and their
 differentials $\delta s_p$ (c) and $\delta s_n$ (d), for the odd-$A$ Eu and 
 Sm isotopes, respectively. }
\label{fig:sp}
\end{center}
\end{figure}

Finally, as yet another clear signature of the QPT, we display in Fig.~\ref{fig:sp} the proton
and neutron separation energies ($s_p$ (a) and $s_n$ (b)) and their
corresponding differentials ($\delta s_p$ (c) and $\delta s_n$ (d))
for the odd-$A$ Eu and Sm isotopes, respectively. 
The separation energies are obtained simply as the difference between the 
eigenvalues of the Hamiltonians ${\hat H}$ and
$\hat H_B$ for the corresponding ground states. Consistent with the results for the other 
characteristic quantities
discussed above, both $\delta s_p$ and $\delta s_n$ exhibit a sharp irregularity at
the transitional nuclei with mass number $A=151$ and 153, for the odd-$A$ Eu and Sm
isotopes, respectively.

\section{Conclusions\label{sec:conclusion}}

A microscopic study of a quantum phase transition
related to the shape of odd-mass nuclei has been carried out using a newly
developed method of Ref.~\cite{nomura2016odd}, based on nuclear
density functional theory (DFT) and the particle-core coupling scheme. 
The deformation energy surface for the even-even core nuclei, and the single-particle
energies and occupation probabilities of the odd fermion (proton or neutron), are obtained from
SCMF calculations based on a choice of the energy density functional and pairing interaction. 
The self-consistent mean-field results determine the parameters of the IBFM
Hamiltonian that is used to calculate spectroscopic properties of odd Eu and Sm
nuclei with mass number $A=147-155$. The corresponding even-even core Sm isotopes   
present one of the best examples of a QPT between spherical and axially-deformed shapes. 
By using this method, characteristic mean-field and spectroscopic
properties that can be related to quantum order parameters of the
QPT have been analysed and, in particular, the differentials of these
quantities that underline the QPT. 
Even though systems with a finite number of particles have been investigated, 
and the control parameter is the integer value of the nucleon
number rather than a continuous parameter, the differentials of several characteristic quantities 
(deformation parameter, q-invariants, excitation energies and separation energies) in the odd-$A$
Eu and Sm nuclei, as well as for the even-even Sm cores, all exhibit a clear discontinuity close to 
$N=90$ which signals the QPT associated with the softness of the collective 
potential in transitional nuclei. 
The results are robust and general, and present a valuable contribution 
towards a systematic study of shape phase transitions in odd-mass nuclei.

\begin{acknowledgements}
 
We would like to thank F. Iachello for valuable discussions. 
K. Nomura acknowledges support from the Japan Society for the Promotion of Science. 
This work has been supported in part by 
the Croatian Science Foundation -- project ``Structure and Dynamics
of Exotic Femtosystems" (IP-2014-09-9159) and the QuantiXLie Centre of Excellence.

\end{acknowledgements}

\bibliography{refs}

\end{document}